%\documentstyle[12pt]{article}
%\setlength{\oddsidemargin}{0mm}
%\setlength{\evensidemargin}{0mm}
%\setlength{\textwidth}{160mm}
%\setlength{\topmargin}{-10mm}
%\setlength{\textheight}{210mm}
%\parindent=10mm
%\parskip=5mm 
 
%\begin{document}
 
%------------------------------------------------------
\parindent 0pt
\magnification\magstep1
\baselineskip 16pt plus 1pt minus 0.5pt
\hsize = 15.0 true cm
\vsize = 22.0 true cm
\abovedisplayskip 16pt plus 5pt minus 6pt
\belowdisplayskip 16pt plus 6pt minus 5pt
\setbox\strutbox = \hbox{\vrule height 13.5 true pt depth
                          6.5 true pt width 0pt} 

%------------------------------------------------------
\hfill 12. Sept. 1991
 
\vskip1.0truecm

\centerline{\bf Isospin Dependent Effective Interaction} 
\centerline{\bf in} 
\centerline{\bf Nucleon-Nucleus Scattering}

\vskip1.5truecm
\centerline{Taksu Cheon }

\centerline{ Department of Physics, Hosei University} 
\centerline{ Chiyoda-ku Fujimi, Tokyo 102, Japan} 
\vskip1.0truecm
\centerline{and } 
 \vskip1.0truecm
\centerline{Kazuo Takayanagi}

\centerline{ Department of Mathematical Sciences, Tokyo Denki University} 
\centerline{ Hatoyama, Saitama 350-03, Japan}
\vskip2.0truecm
\centerline{\bf Abstract}
\vskip1truecm
The isospin-dependent component of the effective nucleon-nucleon 
interaction that causes the $\Delta T=1$  (p, p') and (p, n) reactions off 
nuclei is studied.  It is shown that the  corrections to the impulse approximation
comes from the $g$-matrix type correction and the rearrangement term.    
They are numerically estimated with the isospin-asymmetric nuclear-matter reaction 
matrix approach. 
The analysis of the isobaric analog transitions  ${}^{42}$Ca(p,n)${}^{42}$Sc 
and ${}^{48}$Ca(p,n)${}^{48}$Sc are presented.
\vfill\eject
%-----------------------------------------------------------

\qquad The nucleon direct reaction of few hundred MeV to GeV projectile energy 
has become one of the preferred tools to extract the nuclear structure 
information complementing the reaction with the electromagnetic probes.  Its 
success crucially depends on the good command of the effective 
nucleon-nucleon interaction used in the calculations based on the 
distorted-wave Born approximation (DWBA).  From the theoretical point 
of view, there is a possibility of performing a convergent calculation 
of Watson expansion [1,2] or its variants [3], since the impulse-approximation 
is expected to be a reliable starting point at these energies.  In  
practical analysis with the currently available computational resources, 
however, the treatment based on the density-dependent effective 
interaction [4,5] seems to be the only workable scheme to handle the medium 
corrections. 

\qquad For the elastic scattering, 
the optical model potential obtained by folding the Brueckner reaction matrix ($g$-matrix) [6]
is known to represent the leading  terms in the Watson expansion [7], and  the $g$-matrix
approach has been reasonably successful in describing the
elastic scattering experiments [4,5,8,9].  On the contrary, for the
inelastic  scattering, there had been no such general recipe.   Many DWBA 
calculations had been performed, however, with the density dependent $g$-matrix 
without any theoretical justification.  
In this situation, we have shown  
for the simplest case, i.e. the isoscalar natural parity transitions,  that the
transition potential is the sum of the $g$-matrix and the rearrangement term which is
written in terms of the derivative of the $g$-matrix with respect to 
the nuclear matter density [10,11].    
   Attempts to describe experimental data along this line have 
been pursued in recent years, and met  a great success.  The set of 
density-dependent effective interactions constructed by Kelly and collaborators 
[12-14] has been proven quantitatively accurate and practically useful 
to describe the elastic scattering and isoscalar natural parity excitations.   
In this background, we believe it quite urgent 
to  identify medium corrections in other components of 
the effective nucleon-nucleon interaction than the spin-isospin independent 
part, since one of the advantage of the nucleon projectile over the electron 
is the wider variety of nuclear excitation modes accessible by it.

\qquad	Although the isobaric analog excitation [15] has been one of the 
best-studied nuclear excitation modes, the isospin-dependent part of  
the effective interaction has been given very little theoretical attention. 
This unsatisfactory state of affairs needs to be altered before any systematic 
empirical analysis be performed.  
In this Letter, we intend to identify the  leading medium corrections to 
the isospin dependent component of the  effective nucleon-nucleon interaction
which induces the isovector nuclear  excitations.    
In the formulation of such a process, the 
 $g$-matrix in an asymmetric nuclear matter where the 
Fermi momentum of the proton differs from that of the neutron naturally appears.   
This $g$-matrix is used to evaluate the medium effects on the isovector transitions.
 We report the result of 
 a numerical example at the incident energy  $E_p = 150 MeV$. 
     It is applied
to the isobaric analog transitions ${}^{42}$Ca(p,n)${}^{42}$Sc and
${}^{48}$Ca(p,n)${}^{48}$Sc at $E_p = 135 MeV$.  We hope that this work is to be  the
first step for the microscopic analysis of all the components of effective 
nucleon-nucleon interaction for intermediate energy nucleon scattering.

\qquad	The $g$-matrix in an isospin-asymmetric nuclear matter [16] satisfies the
Bethe-Goldstone equation formally identical to the one in the symmetric nuclear matter:

$$
g = v + v G_0 Q g \quad, \eqno(1)
$$

where $v$ is the free nucleon-nucleon interaction, and  $G_0$ and $Q$ are the 
two-nucleon propagator and the Pauli operator in the medium, respectively.
 We neglect the self-energy term in $G_0$ assuming that the incident proton
has sufficiently high energy.  Then the difference of proton and neutron 
densities, $\rho_p$
and $\rho_n$, generates  {\it the isovector component} of $Q$, in addition to 
the isoscalar component which characterizes the isospin-symmetric nuclear matter.   
 As a result, the effective interaction $g$ of eq.(1) has four independent components
$g_{{\scriptscriptstyle T,T_{z}}}$ specified by the total isospin $T$ and its $z$-component
$T_z$ of the two interacting nucleons.
 We write the $g$-matrix in the following way, featuring the tensor properties in isospin
space: 

$$
g = g^{[0]} + g^{[\tau ]}  {\bf \tau}_1 \cdot {\bf \tau}_2 +
    g^{[\alpha ]} ({\bf \tau}_1 +{\bf \tau}_2) +
    g^{[\beta ]} \sqrt {2/3} 
    [{\bf \tau}_1 \times {\bf \tau }_2]^{(2)} \quad, \eqno(2)
$$

where ${\bf \tau}_1$ and ${\bf \tau}_2$  are the isospin operators of two 
interacting nucleons.
The four coefficients in the eq.(2) are written in terms of  $g_{{\scriptscriptstyle
T,T_{z}}}$ as 

$$\eqalignno{
g^{[0]} &= {1 \over 4} \left( { g_{11} +
   g_{10} + g_{1-1} + g_{00} } \right) \quad, &(3a) \cr
g^{[\tau ]} &= {1 \over {12}} \left( {
   g_{11} + g_{10} + g_{1-1} - 3g_{00} } \right) \quad, &(3b) \cr
g^{[\alpha ]} &= {1 \over 4} \left( {
   g_{11} - g_{1-1} } \right) \quad, &(3c) \cr
g^{[\beta ]} &= {1 \over 6} \left( {
   g_{11} - 2g_{10} + g_{1-1} } \right) \quad. &(3d) \cr
}$$

From eqs.(3a) to (3d), one recognizes that $g^{[0]}$, $g^{[\tau ]}$ and $g^{[\beta ]}$ are
invariant with  respect to the exchange of protons and neutrons in the medium, i.e. $\rho_{p}
\leftrightarrow \rho_n$, while $g^{[\alpha ]}$ changes its sign.  These properties put  
restrictions of the possible form of density dependence of each term.   A transparent view is
obtained by defining the isoscalar density $\rho^{[s]} =$ $\rho_{p} + \rho_n$, and the 
isovector density  $\rho^{[v]} =$ $\rho_{p} - \rho_n$.  Specifically at the isospin-symmetric 
limit $\rho^{[v]} =0$, one obtains

$$ g^{[\alpha]} = g^{[\beta]} =
{{\partial  } \over { \partial \rho ^{[v]}}}g^{[0]} =
   {{\partial  } \over { \partial \rho ^{[v]}}}g^{[\tau ]} =
   {{\partial  } \over { \partial \rho ^{[v]}}}g^{[\beta ]} = 0
 \quad. \eqno(4a)
$$

Also, taking the leading density-dependence of $g^{[\alpha]}$, one obtains

$$
g^{[\alpha ]} \approx  
  \rho ^{[v]}{{\partial  } \over { \partial \rho ^{[v]}}}g^{[\alpha ]}
 \quad. \eqno(4b)
$$

\qquad   We now consider the elastic and inelastic scatterings in order.    
The  elastic optical model potential $U$ of a nucleon in the isospin 
asymmetric nuclear matter is given by
folding the effective interaction $g$ with the asymmetric nuclear density.
 It is convenient to split the resultant optical  model potential into isoscalar and 
isovector part  {\it \'a la}
Lane as [17]

$$
U = U^{[s]} + U^{[v]} \, {\bf \tau }_z \quad, \eqno(5)
$$

where each term is given by

$$\eqalign{
U^{[s]} & = g^{[0]} \rho ^{[s]} + g^{[\alpha ]} \rho ^{[v]} \cr
         & \approx  g^{[0]} \rho ^{[s]}
 \quad, }
\eqno(6a)
$$

and

$$ \eqalign{
U^{[v]} & = g^{[\alpha ]} \rho ^{[s]} +
   \left( g^{[\tau ]} + g^{[\beta ]} \right) \rho ^{[v]} \cr
 & \approx  g^{[\alpha ]} \rho ^{[s]} +
    g^{[\tau ]} \rho ^{[v]} \cr
 & \approx \left(
  \rho ^{[s]}{{\partial  } \over { \partial \rho ^{[v]}}}g^{[\alpha ]}
 + g^{[\tau]} \right) \rho^{[v]}
 \quad.  }
\eqno(6b)
$$

 In the derivation of $U^{[s]}$ and $U^{[v]}$, we have dropped all terms that
are of second or higher orders in powers of $\rho^{[v]}$.   In the last line of
eq.(6b), the relation (4b) is used.
 The above expression shows that the optical model potential $U$ of eq.(5) 
is obtained by folding

  $$ \eqalign{
   v_{el} &=  g^{[0]} \rho ^{[s]}
 + \bigl(
  \rho ^{[s]}{{\partial  } \over { \partial \rho ^{[v]}}}g^{[\alpha ]}
 + g^{[\tau]} \bigr)  {\bf \tau}_1 \cdot {\bf \tau}_2 \cr
 & \equiv v_{el}^{0} + v_{el}^{\tau} \,\, {\bf \tau}_1 \cdot {\bf \tau}_2 
 \quad, } \eqno(7)
  $$
 with the nuclear densities $\rho^{[s]}$ and $\rho^{[v]}$.    
One can regard $v_{el}$ {\it defined by} eq.(7) as the effective interaction  
for the elastic scattering.

\qquad  For the inelastic scatterings, both through the
diagrammatical analysis of Watson expansion and the macroscopic collective
excitation model, one can relate the inelastic transition potential and the density
derivative of the optical potential $U$ for the elastic scattering [10,11].    
Assuming that the transition of the target nucleus is described by 
the isoscalar and isovector transition densities  $\rho_{tr}^{[v]}$ and
$\rho_{tr}^{[v]}$, 
one can write the transition potential $U_{tr}$ which describe the inelastic 
scattering as

$$
U_{tr}= 
 \rho_{tr} ^{[s]}{{\partial  } \over { \partial \rho ^{[s]}}} U^{[s]}
+  \rho_{tr} ^{[v]}{{\partial  } \over { \partial \rho ^{[v]}}} U^{[v]} \, \tau_{z} 
 \quad. \eqno(8)
$$ 

This yields, for $N=Z$ nuclei ($\rho^{[v]}=0$), the expression

$$
U_{tr}= \rho^{[s]}_{tr} (  g^{[0]} +\rho ^{[s]}{\partial  \over 
{ \partial \rho ^{[s]}}}g^{[0]} )
 + \rho^{[v]}_{tr} 
(  g^{[\tau]} + \rho ^{[s]}{\partial  \over 
{ \partial \rho ^{[v]}}}g^{[\alpha]}
)\tau_{z} \quad, \eqno(9)
$$

which explictly shows that the transition potential can be obtained from
the effective interaction

$$
v_{in} =  v_{in}^{[0]} + v_{in}^{[\tau ]} 
 {\bf \tau }_1\cdot {\bf \tau} _2  
 \quad, \eqno(10)
$$

where the isospin-independent  and 
dependent components,  $ v_{in}^{[0]}$ and $v_{in}^{[\tau ]}$, are given  by

$$
v_{in}^{[0]} =  g^{[0]} + \rho ^{[s]}{\partial  \over { \partial \rho ^{[s]}}}g^{[0]}
\quad, \eqno(11a)
$$
 
and 

$$
v_{in}^{[\tau ]} = g^{[\tau ]}\ +
 \rho ^{[s]}{\partial  \over { \partial \rho ^{[v]}}}g^{[\alpha ]}
\quad.  \eqno(11b) 
$$

  The second term of eq.(11a) is the so-called 
rearrangement term for the isoscalar transition [10] which roughly doubles the density
dependence of the inelastic effective interaction.
 Eq.(11b) is one of the main result of this Letter.  Its second term is analogous to the one
in eq.(11a), and  no less remarkable since it persists at $\rho^{[v]}=0$ 
despite that  $g^{[\alpha]}$  itself vanishes at this limit.
 We now compare  eq.(7) with eq.(11), i.e.  the effective interactions for elastic and 
inelastic scatterings.    For the isospin-independent component, one recognizes

$$
v_{in}^{[0]} =\left( {1 +\rho ^{[s]}
{\partial  \over { \partial \rho ^{[s]}}}} \right) v_{el}^{[0]} \quad, \eqno(12)
$$

 which has been instrumental in determining the empirical density-dependent 
interaction [12-14].  
For the isospin-dependent component, one finds

$$
v_{in}^{[\tau ]} =v_{el}^{[\tau ]}  = g^{[\tau ]}\ +
 \rho ^{[s]}{\partial  \over { \partial \rho ^{[v]}}}g^{[\alpha ]} \quad. \eqno(13)
$$
  
This shows that 
the extra factor appearing in eq.(12)
can be formally eliminated in the case of isospin-dependent component.  
We stress that this {\it does not} mean the absence of the rearrangement term.  
It is achieved by  natural but  purely formal redefinition, eq.(7).  
In principle, therefore, with eq.(13), the 
isospin-dependent empirical interactions can be constructed from 
the combined analysis of the elastic scattering and the $\Delta T = 1$ 
inelastic and charge-exchange reactions on $N \sim Z$ nuclei.

\qquad We now turn to the numerical assessment of the medium 
correction to the isospin-dependent component of the effective 
interaction.        The detailed description of the 
technique to solve eq.(1) in an isospin-asymmetric nuclear matter and the full
 results will be published elsewhere.  Here, we show the 
relevant result of an example of such calculation.  At the projectile energy of 
$E_p = 150 MeV$, Reid Soft Core potential as the input free two nucleon potential, 
we obtain the following number for the volume integrals of isospin dependent 
components of $g$-matrix.

$$\eqalignno{
  g^{[\tau]} & = [\{22 + 75i\}+\{-24 - 53i\}\rho^{[s]}] 
            + [\{8 + 22i\}+\{17 - 17i\}\rho^{[s]}] 
            {\bf L} \cdot {\bf S}  \,\,  ,	&(14) \cr
 g^{[\alpha]} &= [\{6+38i\}\rho^{[v]}] 
         + [\{-1+1i\}\rho^{[v]}]  {\bf L} \cdot {\bf S}  \quad, &(15) \cr
}$$

 where the densities  $\rho^{[s]}$ and  $\rho^{[v]}$ are normalized to the nuclear matter 
saturation density.
The isospin dependent component of the $g$-matrix, $g^{[\tau]}$, is identical to the one
in the symmetric nuclear matter in ref.[18]. It includes the density dependence induced
by the nonlocality.  Only the contribution from the allowed channel is included, and thereby
the antisymmetrization is taken into account. 
 The numbers in eqs.(14) and (15) can be also interpreted as the strength of the 
interaction in zero-range approxiamation.   One observes that, at this energy,  
the density dependence of  $g^{[\alpha]}$ 
tends to cancel that of $g^{[\tau]}$,  which already exists in 
the  $g$-matrix in the symmetric nuclear matter, in the  calculation of the isospin-dependent
effective  interaction, eq.(13), especially in the imaginary part of the central force.   
More to the point, these two density-dependent  corrections are of the same order numerically.  
It clearly shows that the partial inclusion  of the medium correction, only with $g$-matrix, 
for example, cannot be very meaningful.

\qquad In order to show how the density-dependent 
effective interaction eqs.(13-14) works in the
inelastic scattering, we show the result of DWBA calculations [19] for the isobaric analog 
transitions ${}^{42}$Ca(p,n)${}^{42}$Sc and  ${}^{48}$Ca(p,n)${}^{48}$Sc at $E_p = 135
MeV$.   For the facility of the calculation, we adopt following two approximations: 
First,  we take the  empirical optical model potential   by Schwandt  {\it et al} [20], 
instead of calculating it microscopically.   
Second, we replace the density-independent part of eqs.(14) and (15) by
the three-range Yukawa parametrization of free two-nucleon  scattering matrix by Franey and
Love [21] at $E_p = 140 MeV$ (FL140).
  The density-dependent portion of the above effective interactions is treated as Yukawa 
force of very short range.
   The calculated differential cross section is shown in Figures, 
where dashed line corresponds to the impulse approximation with FL140 interaction; 
dotted-dashed line, the result with $v_{in}^{[\tau]}$ of eq.(11b) being replaced by $g^{[\tau]}$;  
solid line, the result with the full medium correction included.  We can see the 
following effect: First, the calculation with the $t$-matrix overestimates the experimental
cross section [22].  Second,  by changing the interaction  to the $g$-matrix, 
the calculation lowers the cross section, but overshoot the data.   
Last, by including all the medium effects, the calculation hits the experimental data points 
very well.    
A warning is due, however, to  the overemphasis on the favourable comparison with the
experiment at this point,  since the uncertainty of the isospin-dependent component of the 
free $t$-matrix is  already very large.   
Also, the approximations in the current  calculation leaves some space for further improvement.  
Rather, we would  reiterate our basic point that the medium modification should be, and can be  
estimated with the proper inclusion of the rearrangement term.

\vskip1.0truecm

\indent We acknowledge our gratitude to Prof. K. Yazaki for the valuable discussions.  
Thanks are also due to Prof. H. Sakai for his advice in experimental and 
computational aspect of this work.  The numerical work has been performed on 
SPARCStation2 at Hosei University, Department of Physics, and on VAX 6000/440 
of Meson Science Laboratory of University of Tokyo.   
This work is financially supported by Research Institute for Technology, Tokyo Denki University.

\vfill\eject

{\bf References}
\vskip1.0truecm

  \item{1.}	{K. M. Watson, Rev. Mod. Phys. 30, 565 (1958).}

  \item{2.}	{G. Goldberger and K. M. Watson, Collision Theory (John Wiley, N. Y., 1964).}

  \item{3.}	{A. Kerman, H. McManus and R. Thaler, Ann. Phys. (N.Y.) 8, 551 (1959).}

  \item{4.}	{J. P. Jeukenne, A. Lejeune and C. Mahaux, Phys. Rev. C10, 1391 (1974); 
  ibid, C16, 80 (1977).}

  \item{5.}	{F. A. Brieva and J. R. Rook, Nucl. Phys. A291, 299 (1977); ibid, A291,  317
(1977).}

  \item{6.}	{K. Brueckner, R. J. Eden and N. C. Francis, Phys. Rev. 100, 891 (1955).}

  \item{7.}	{J. H\"ufner and C. Mahaux, Ann. Phys. (N.Y.) 73, 525 (1972).}

  \item{8.}	{H. V. von Geramb, L. Rikus and K. Nakano, Proc. RCNP Int. Symp. on Light 
  Ion Reaction Mechanism, Osaka, Japan (1983) p.78. }

  \item{9.}	{L. Ray, Phys. Rev. C41, 2816 (1990).}

  \item{10.}	{T. Cheon, K. Takayanagi and K. Yazaki, Nucl. Phys. A445, 227 (1985).}

  \item{11.}	{T. Cheon, K. Takayanagi and K. Yazaki, Nucl. Phys. A437, 301 (1985).}

  \item{12.}	{J. J. Kelly, Phys. Rev. C39, 2020 (1989).}

  \item{13.}	{J. J. Kelly, J. M. Finn, W. Bertozzi, T. N. Buti, F. W. Hersman,C. E. 
 Hyde-Wright, M. V. Hynes, M. A. Kovash, B. Murdock, P. Ulmer, A. D. Bacher, G.
 T. Emery, C. C. Foster, W. P. Jones, D. W. Miller, and B. L. Berman, 
 Phys. 	Rev. C41, 2504 (1990).}

  \item{14.}	{B. S. Flanders, J. J. Kelly, H. Seifert, D. Lopiano, B. Aas, A. 
 Azizi, G. Igo, 	G. Weston, C. Whitten, A. Wong, M. V. Hynes, J. McClelland,
 W. Bertozzi, J. 	M. Finn, C. E. Hyde-Wright, R. W. Lourie, B. E. Norum,
 P. Ulmer, B. L. 	Berman, Phys. Rev. C43, 2103 (1990).}

  \item{15.}	{J. D. Anderson and C. Wong, Phys. Rev. Lett. 7, 250 (1961).}

  \item{16.}	{S. Nishizaki, T. Takatsuka, N. Yahagi, and J. Hiura, Iwate University 
 Preprint 	HSIU-20 (1991).}

  \item{17.}	{A. M. Lane, Nucl. Phys. 35, 676 (1962).}

  \item{18.}	{T. Cheon and E. F. Redish, Phys. Rev. C39, 1173 (1989).}

  \item{19.}	{R. Shaeffer and J. Raynal, Program DWBA70 (unpublished); extended version 
 DW82 by J. R. Comfort (unpublished).}

  \item{20.}	{P. Schwandt, H. O. Meyer, W. W. Jacobs, A. D. Bacher, S. E. Vigdor, and M. 
 D. Kaitchuck, Phys. Rev. C26, 55 (1982).}

  \item{21.}{M. A. Franey and W. G. Love, Phys. Rev. C31, 488 (1985).}

  \item{22.}	{B. D. Anderson, M. Mostajabodda'vati, C. Lebo, R. J. McCarthy, L. Garcia,
 J. 	W. Watson, and R. Madey, Phys. Rev. C43, 1630 (1991).}

\vfill\eject

{\bf Figure Captions}

\vskip1.0truecm
Fig. [I]:  The differential cross section for the reaction ${}^{42}$Ca(p,n)${}^{42}$Sc.  
The dashed line is the result of the impulse approximation.  
The dotted-dashed curve  partially accounts for the medium corrections, while the 
solid line includes all the medium modifications.

\vskip0.5truecm
Fig. [II]: The differential cross section for the reaction ${}^{48}$Ca(p,n)${}^{48}$Sc.
   Notation is
the same as for fig.[I].

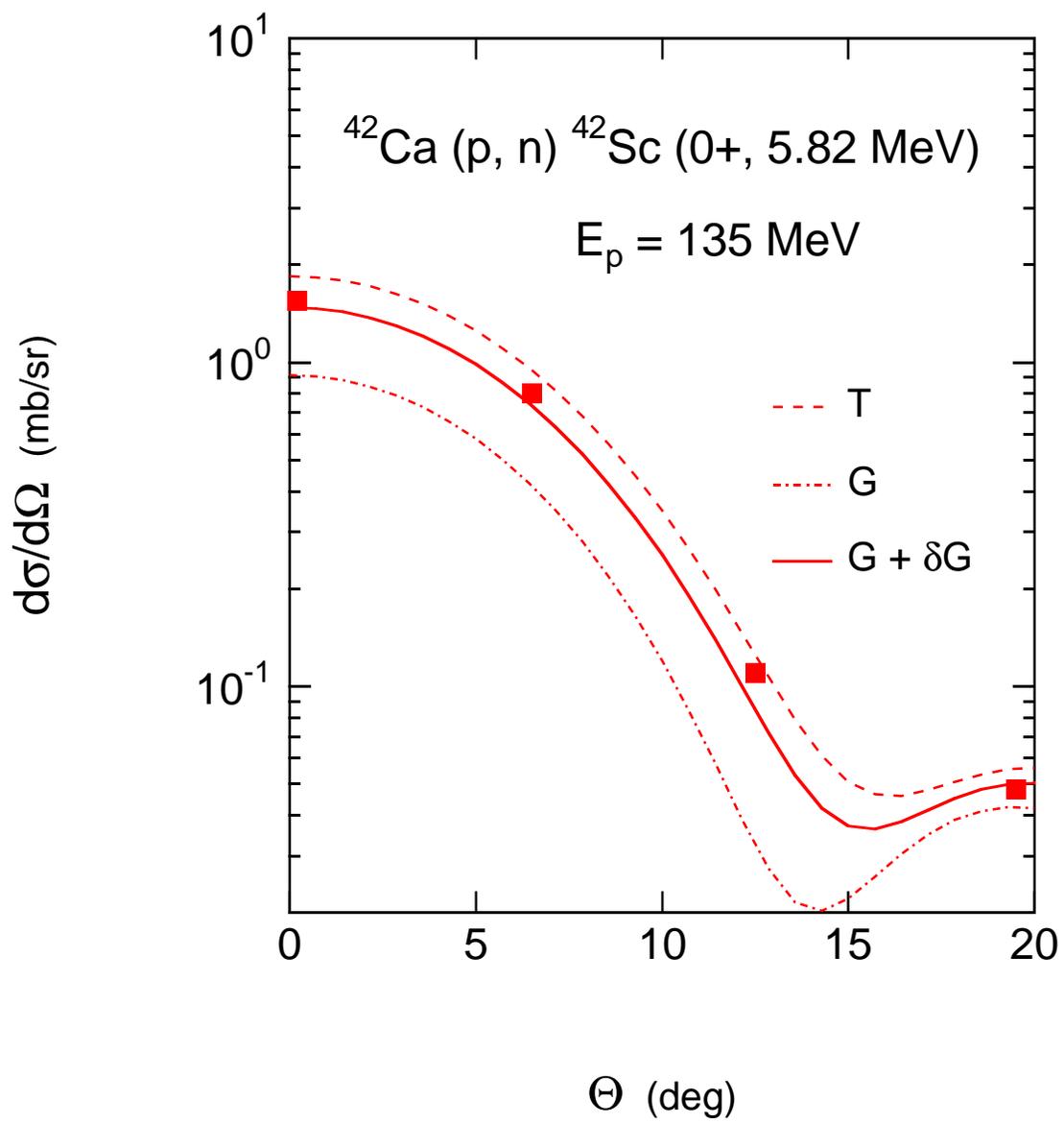
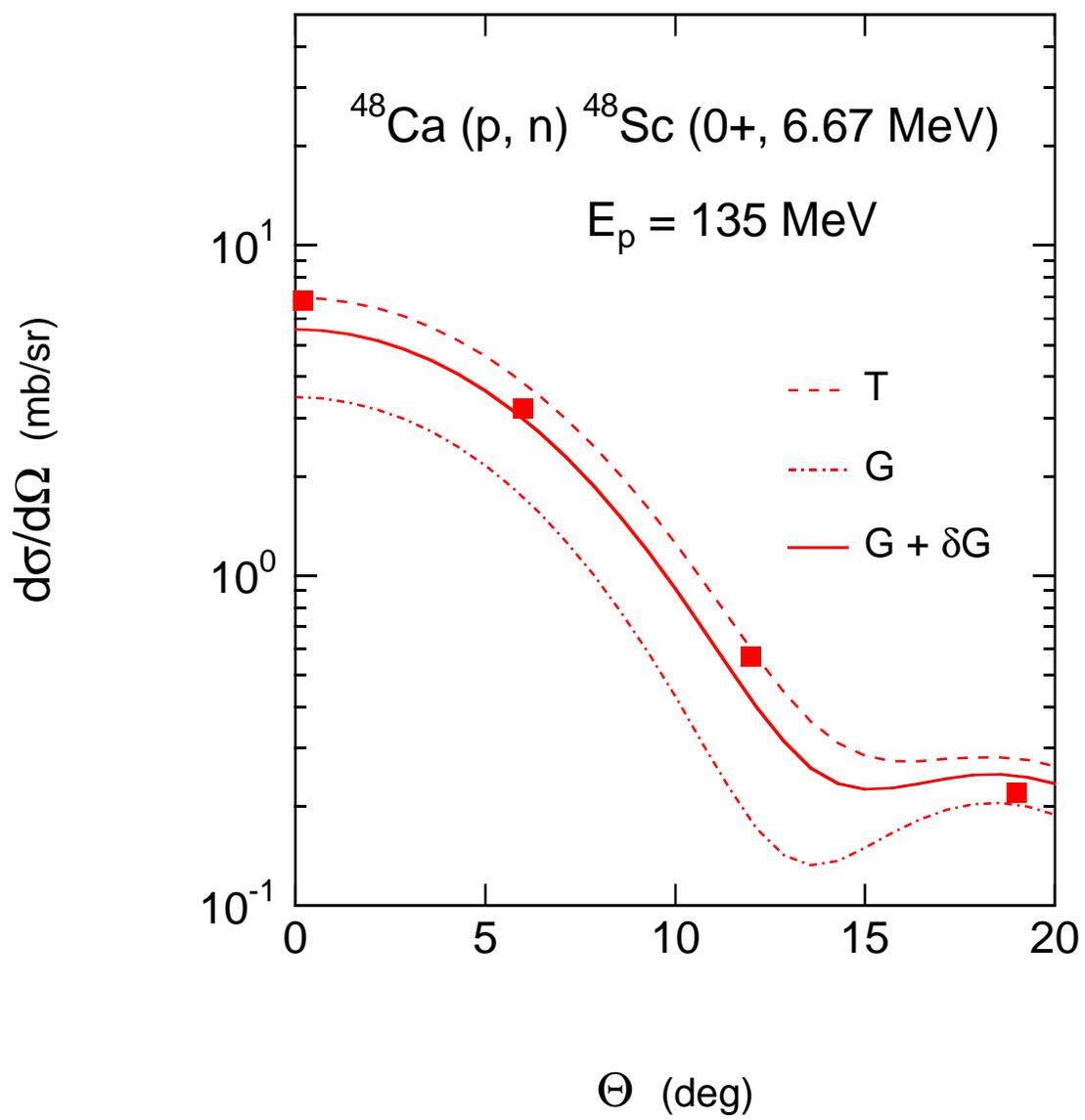
\bye